\documentclass[fleqn,12pt,twoside]{article}
\usepackage{espcrc1}

\usepackage{graphicx}
\usepackage[figuresright]{rotating}

\newcommand{\AmS}{{\protect\the\textfont2
  A\kern-.1667em\lower.5ex\hbox{M}\kern-.125emS}}

\title{ Measuring the $\phi$ meson width in the medium from $p$ induced
$\phi$ production in nuclei}

\author{V.K.~Magas\address{Departamento de F\'{\i}sica Te\'orica and IFIC
 Centro Mixto Universidad de Valencia-CSIC\\
 Institutos de Investigaci\'on de Paterna, Apdo. correos 22085,
 46071, Valencia, Spain}, 
 L.~Roca\addressmark, E.~Oset\addressmark
 }
       
\begin{document}

\maketitle

\begin{abstract}
We study the  $A$
dependence of the $\phi$ meson production cross section in proton nucleus
reactios at energies just above threshold, which are accessible in an
experimental facility like COSY.  
This $A$ dependence has two sources: the
distortion of the incident proton and the  absorption of the 
$\phi$ in its way out of the nucleus. This second 
process reduces the cross section in about a factor two in heavy
nuclei. Thus we show that the $A$ dependence of the cross section 
contains valuable information on the $\phi$ width in the 
nuclear medium.
\vspace{1pc}
\end{abstract}

The study of the properties of vector mesons in a nuclear
medium is one of the subjects in hadron physics which receives
continuous attention (see for instance 
Ref.~\cite{Rapp:2000ej}).  Although originally the $\rho$ meson
properties were mostly investigated, nowadays the $\phi$ properties 
have got a lot of
interest, because the medium renormalization in this case
is more drastic than that of the
$\rho$. Indeed, predictions of an increase of the $\phi$ width
by a factor five or six \cite{Oset:2001eg,dani} to ten
\cite{Klingl:1998tm}, at normal nuclear matter density,
have been made using different chiral
approaches. Different reactions have been
studied or suggested to test experimentally 
this large width 
\cite{Pal:2002aw,Yokkaichi:wn,Klingl:1998tm,Oset:2001na,Mosel,daniluis}.

The aim of the present work is to propose a new method to determine
the $\phi$ width in the nuclear medium \cite{MRO}. The traditional method
in the works quoted above (except \cite{daniluis}) is to look
for a broadening of the $\phi$ width reconstructed from the
invariant mass of its decay products. Here, instead, we use a
different philosophy and we investigate the $A$ dependence of
$\phi$ production in $pA$ collisions, in a similar  way as it was
done in \cite{daniluis}  with the $\phi$ photoproduction in
nuclei,  which is the subject of experimental investigation at
Spring8/Osaka \cite{imai}.  The advantage of performing the
reaction slightly above threshold is that one can rule out the
contribution from coherent $\phi$ production which might obscure
the interpretation of the experimental results in \cite{imai}. 
The present reaction, with its particular kinematics, is
amenable of experimental performance at facilities like COSY.

In order to implement the relevant nuclear effects in the $\phi$
production cross section we will use a model based on many body
techniques, successfully applied in the past in many works
\cite{Salcedo:md,Carrasco:vq} to study the interaction of
different particles with nuclei. The model assumes a local Fermi
sea at each point in the nucleus and provides a very simple and
accurate way to account for the Fermi motion of the initial
nucleon and the Pauli blocking of the final ones. On the other
hand, we have to take into account the distortion of the incoming
nucleon and the final $\phi$ meson in the their way through the
nucleus, which are evaluated in the present work  using an
eikonal approximation (see \cite{MRO} for more details).

We also assume that the $T$ matrix for our process is angular independent. This
is supported by the experiment \cite{Balestra:2000ex} 
where the angular dependence of 
$pp\to pp\phi$ is almost flat.

For the evaluation of ${\cal I}m\Pi$, imaginary part of the $\phi$ 
selfenergy in nuclear matter, we use
the results of the model of Ref.~\cite{dani} and its extension to
finite $\phi$-meson momentum done in \cite{daniluis}. This model
is  based on the modification of the $\bar K K$ decay channel in
the medium by means of a careful treatment of the in medium
antikaon selfenergies \cite{Oset:2001eg}. It uses a
selfconsistent coupled channel unitary calculation, based on
effective chiral Lagrangians, and taking into account Pauli
blocking, pion selfenergies and  mean-field potentials of the
baryons (for the S-wave part) and hyperon-hole excitations (for
the P-wave part).
It corresponds to
a $\phi$ medium width at rest  at $\rho=\rho_0$ of the order of
$24\ MeV$.

We also take into account
$\phi$ production from two-step processes. Let us imagine we
have a $pN$ collision of the initial proton going to any other
channel than $\phi$ production. In such cases the fast incoming
proton will usually survive although with a reduced energy, by
means of which it still can contribute to $\phi$ production. We
estimate the contribution from this mechanism, based on the 
$p$ energy loss in the first collision $\Delta E \simeq 400\ MeV$ or 
more \cite{MRO}.

We also study two-step process with $\Delta$ intermediate states: 
$NN\rightarrow N \Delta$ reaction 
followed by
$\Delta N\rightarrow NN \phi$. 
This two body process would benefit with respect to the one considered above from the fact that 
the $\Delta$  couples more strongly to pions and
vectors than the nucleon. For instance, one can consider a mechanism from the model of 
\cite{titov,barz}  which, with respect to the same one with a
nucleon instead of a $\Delta$, would benefit from the factor $f_{\pi N\Delta}/f_{\pi NN}=2.13$ 
in the amplitude, hence a factor 4.5 in the cross section. Not surprisingly our results show
that the two-step mechanism with $\Delta$ excitation is
more relevant than the two-step mechanism involving only
nucleons.
We shall also  distinguish between $\Delta$ excitation
on the target and $\Delta$ excitation on the projectile. 
The mechanism of $\Delta$ excitation in
the projectile appears to be more important than that of $\Delta$
excitation in the target in the present reaction.

We performed calculations for the
following nuclei:   ${}^{12}_6C$, ${}^{16}_{8}O$, 
${}^{24}_{12}Mg$,  ${}^{27}_{13}Al$, ${}^{28}_{14}Si$, 
${}^{31}_{15}P$,   ${}^{32}_{16}S$,  ${}^{40}_{20}Ca$,  
${}^{56}_{26}Fe$, ${}^{64}_{29}Cu$,  ${}^{89}_{39}Y$,  
${}^{110}_{48}Cd$,  ${}^{152}_{62}Sm$,  ${}^{208}_{82}Pb$,
${}^{238}_{92}U$. 
 
\begin{figure}[htb]
\begin{center}
\includegraphics[width=12cm]{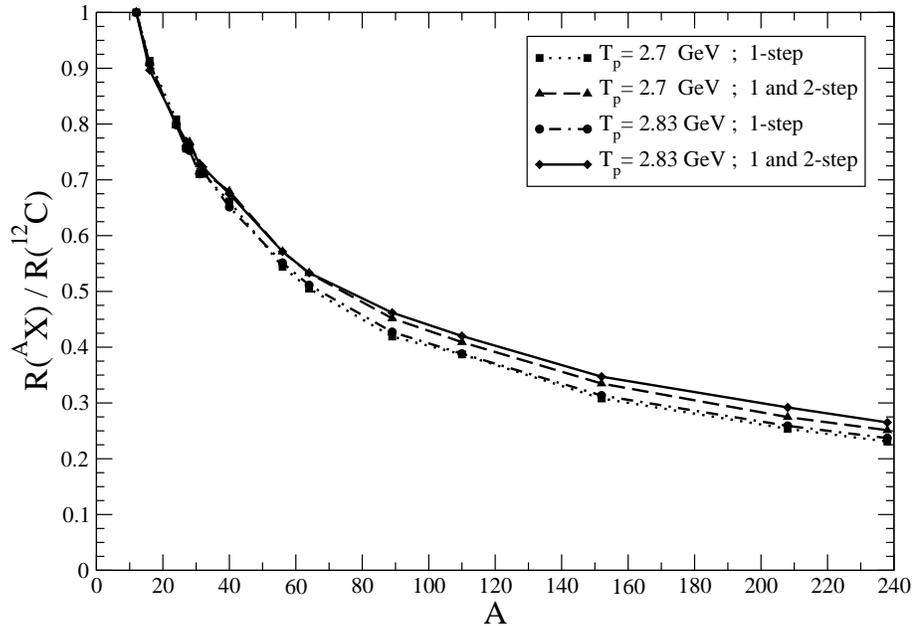}
\end{center}
\vspace{-1.5cm}
\caption{Ratio of the nuclear cross section normalized to
$^{12}C$ for two different incident proton kinetic energies,
$T_p$, including or not 
 the two-step mechanisms.
  The two-step process with nucleon intermediate states
 was evaluated with $\Delta
E=400\ MeV$. From \cite{MRO}.} 
\label{ff6}
\end{figure}

In our analyses we are concerned about the $A$
dependence, no so much on the absolute values of the cross
sections, since the $\phi$ absorption effect is reflected in
this $A$ dependence.  
 To see this most clearly, we calculate the following observable - the normalized ratio: 
$R(^A X)/R(^{12}C)$, where $R(^A X)=\sigma_{A}/(A\sigma_{free})$. 
Fig.~\ref{ff6} shows 
corresponding curves for the one-step
and one- plus two-step mechanisms and for different energies.
 We see that
this normalized $R$ changes very little when including the two-step
mechanisms for both the $T_p$ considered.
Note that for the energy closer to the threshold
($T_p=2.7\textrm{ GeV}$) the changes due to the two-step
contributions are smaller.

Thus we conclude that the $A$ dependence obtained
in the present work is  reliable and the calculations
clearly show that proton induced $\phi$ production in nuclei at
energies just above threshold can indeed be used to get
information on the $\phi$ width in the medium.

\begin{figure}[htb]
\vspace{-1.0cm}
\begin{center}
\includegraphics[width=13cm,angle=-90]{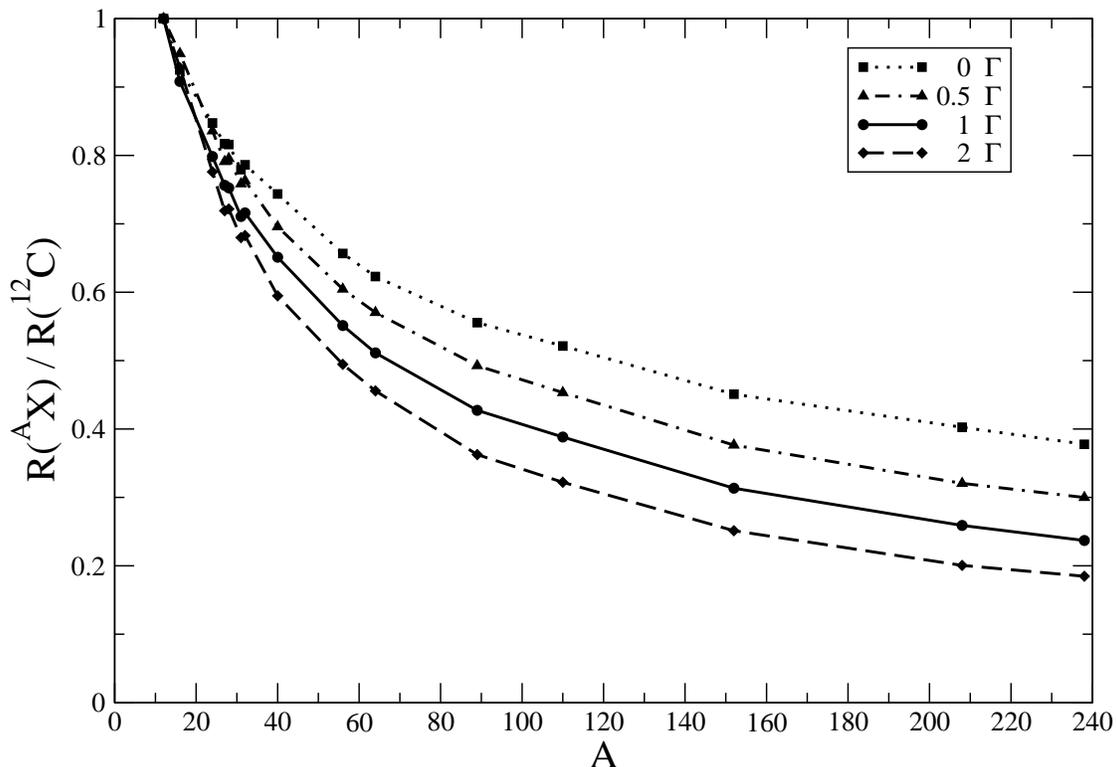}
\end{center}
\vspace{-2.5cm}
\caption{Ratio of the nuclear cross section normalized to
$^{12}C$ for $T_p=2.83\textrm{ GeV}$ and multiplying the $\phi$ width
in the medium, $\Gamma$, by different factors. From \cite{MRO}.} 
\label{ff7}
\end{figure}

In order to see which experimental precision is needed to
get a definite information on the $\phi$ width in the medium, we
have performed the same calculations assuming $\phi$ widths in
the medium to be one half or twice the width used so far
\cite{dani,daniluis}.
In Fig.~\ref{ff7} we show the results of these
calculations for $T_p=2.83\textrm{ GeV}$ (without the inclusion of the
two-step processes).
Comparing Figs.~\ref{ff6} and \ref{ff7} we clearly see that the
uncertainties due to the two-step mechanism are far smaller
than the differences in the results obtained by using these
different $\phi$ widths. The three curves shown there  should serve
to get a fair answer about the $\phi$ width in the medium by
comparing with experimental results.
The uncertainties one might have from the approximate
knowledge of the two-step processes still would allow us to be
sensitive to the value of the $\phi$ width in the medium to the
level of $25$\% of the $\phi$ width we have used.

Our calculations 
were done in symmetric nuclear matter, i.e. in
order to calculate $R=\sigma_{A}/(A\sigma_{free})$ we
took a total free elementary $\phi$ production cross section
$\sigma_{free}=(\sigma_{pn,\phi}+\sigma_{pp,\phi})/2$.
Experimentally we have poor knowledge about these
elementary cross sections 
\cite{Balestra:2000ex}. 
Nevertheless, our results can still be used to
compare with experiment for asymmetric nuclei if
one takes for $\sigma_{free}$ the isospin weighted combination  
$(N\sigma_{pn,\phi}+Z\sigma_{pp,\phi})/A$.
Hence, in order to extract the optimum information
on the $\phi$ width it would be useful to have data on $\phi$
production on neutron targets, for what experiments on the
deuteron would also be most welcome.

{\bf Acknowledgments \ }
One of us, L.R., acknowledges support from the
Ministerio de Educaci\'on, Cultura y Deporte.
This work is partly supported by DGICYT contract number BFM2003-00856,
and the E.U. EURIDICE network contract no. HPRN-CT-2002-00311.


\begin{thebibliography}{99}

\bibitem{Rapp:2000ej}
R.~Rapp and J.~Wambach,
Adv.\ Nucl.\ Phys.\  25 (2000) 1.

\bibitem{Oset:2001eg}
E.~Oset and A.~Ramos,
Nucl.\ Phys.\ A 679 (2001) 616.

\bibitem{dani}
D.~Cabrera and M.~J.~Vicente Vacas,
Phys.\ Rev.\ C  67 (2003) 045203.

\bibitem{Klingl:1998tm}
F.~Klingl, T.~Waas and W.~Weise,
Phys.\ Lett.\ B  431 (1998) 254.

\bibitem{Pal:2002aw}
S.~Pal, C.~M.~Ko and Z.~w.~Lin,
Nucl.\ Phys.\ A  707 (2002) 525.

\bibitem{Yokkaichi:wn}
S.~Yokkaichi  et al.  [KEK-PS-E325 Collaboration],
Nucl.\ Phys.\ A  638 (1998) 435.

\bibitem{Oset:2001na}
E.~Oset, M.~J.~Vicente Vacas, H.~Toki and A.~Ramos,
Phys.\ Lett.\ B  508 (2001) 237.

\bibitem{Mosel}
P.~Muhlich, T.~Falter, C.~Greiner, J.~Lehr, M.~Post and U.~Mosel,
Phys.\ Rev.\ C  67 (2003) 024605.

\bibitem{daniluis}
D.~Cabrera, L.~Roca, E.~Oset, H.~Toki and M.~J.~V.~Vacas,
Nucl.\ Phys.\ A 733 (2004) 130.


\bibitem{MRO}
V.~K.~Magas, L.~Roca and E.~Oset,
nucl-th/0403067.



\bibitem{imai}
J.~K.~Ahn {\it et al.},
nucl-ex/0411016.

\bibitem{Salcedo:md}
L.~L.~Salcedo, E.~Oset, M.~J.~Vicente-Vacas and C.~Garcia-Recio,
Nucl.\ Phys.\ A  484 (1988) 557.

\bibitem{Carrasco:vq}
R.~C.~Carrasco and E.~Oset,
Nucl.\ Phys.\ A  536 (1992) 445.


\bibitem{Balestra:2000ex}
F.~Balestra  et al.  [DISTO Collaboration],
Phys.\ Rev.\ C  63 (2001) 024004.




\bibitem{titov}
A.~I.~Titov, B.~Kampfer and B.~L.~Reznik,
Eur.\ Phys.\ J.\ A  7 (2000) 543.

\bibitem{barz}
H.~W.~Barz and B.~Kampfer,
Nucl.\ Phys.\ A 683 (2001) 594.

\end{thebibliography}
\end{document}